\documentclass[a4paper]{jpconf}

\usepackage{graphicx,natbib,txfonts}

\bibpunct{[}{]}{;}{a}{}{,}

\begin{document}

\newcommand{\aj}{AJ}%
\newcommand{\araa}{ARA\&A}%
\newcommand{\apj}{ApJ}%
\newcommand{\apjl}{ApJ}%
\newcommand{\apjs}{ApJS}%
\newcommand{\ao}{Appl.~Opt.}%
\newcommand{\apss}{Ap\&SS}%
\newcommand{\aap}{A\&A}%
\newcommand{\aapr}{A\&A~Rev.}%
\newcommand{\aaps}{A\&AS}%
\newcommand{\azh}{AZh}%
\newcommand{\baas}{BAAS}%
\newcommand{\jrasc}{JRASC}%
\newcommand{\memras}{MmRAS}%
\newcommand{\mnras}{MNRAS}%
\newcommand{\pra}{Phys.~Rev.~A}%
\newcommand{\prb}{Phys.~Rev.~B}%
\newcommand{\prc}{Phys.~Rev.~C}%
\newcommand{\prd}{Phys.~Rev.~D}%
\newcommand{\pre}{Phys.~Rev.~E}%
\newcommand{\prl}{Phys.~Rev.~Lett.}%
\newcommand{\pasp}{PASP}%
\newcommand{\pasj}{PASJ}%
\newcommand{\qjras}{QJRAS}%
\newcommand{\skytel}{S\&T}%
\newcommand{\solphys}{Sol.~Phys.}%
\newcommand{\sovast}{Soviet~Ast.}%
\newcommand{\ssr}{Space~Sci.~Rev.}%
\newcommand{\zap}{ZAp}%
\newcommand{\nat}{Nature}%
\newcommand{\iaucirc}{IAU~Circ.}%
\newcommand{\aplett}{Astrophys.~Lett.}%
\newcommand{\apspr}{Astrophys.~Space~Phys.~Res.}%
\newcommand{\bain}{Bull.~Astron.~Inst.~Netherlands}%
\newcommand{\fcp}{Fund.~Cosmic~Phys.}%
\newcommand{\gca}{Geochim.~Cosmochim.~Acta}%
\newcommand{\grl}{Geophys.~Res.~Lett.}%
\newcommand{\jcp}{J.~Chem.~Phys.}%
\newcommand{\jgr}{J.~Geophys.~Res.}%
\newcommand{\jqsrt}{J.~Quant.~Spec.~Radiat.~Transf.}%
\newcommand{\memsai}{Mem.~Soc.~Astron.~Italiana}%
\newcommand{\nphysa}{Nucl.~Phys.~A}%
\newcommand{\physrep}{Phys.~Rep.}%
\newcommand{\physscr}{Phys.~Scr}%
\newcommand{\planss}{Planet.~Space~Sci.}%
\newcommand{\procspie}{Proc.~SPIE}%

\title{10 $\mu$m interferometry of disks around young stars}

\author{Roy van Boekel}

\address{Max-Planck-Institut f\"ur Astronomie, K\"onigstuhl~17, D-69117
  Heidelberg, Germany}

\ead{boekel@mpia.de}

\begin{abstract}
  
  This contribution reviews results from interferometric observations of
  circumstellar disks around young stars of $\lesssim$3\,M$_{\odot}$,
  performed with the MIDI instrument operating in the 10\,$\mu$m spectral
  region. Two main topics, the disk structure on $\sim$1-10\,AU scales and the
  dust properties in the same region, are illustrated with several examples
  of MIDI studies, covering various evolutionary stages. The spatially
  resolved observations largely confirm SED-only based hypotheses on disk
  structure, yet also reveal degeneracies that may occur in such SED
  modeling. The properties of the dust on the disk surface show a strong
  radial dependence: the dust close to the central star has generally larger
  grain sizes and in particular a much higher crystallinity than the dust in
  more remote disk regions.
\end{abstract}

\section{Introduction}
Circumstellar disks are ubiquitous around stars during much of their
formation. Disks have been studied for numerous years around young stars of
$\sim$1 and 2-3 solar masses, known as "T~Tauri" and "Herbig~Ae" (HAe) stars,
respectively. Excellent overviews can be found in the "Protostars and Planets
V" book \cite{2007prpl.conf.....R}. More recently, evidence has been found for
disks around brown dwarfs \cite{2005Sci...310..834A} as well as high mass
stars \cite{2005Natur.437..112J,2005Natur.437..109P}, thus extending the mass
range of young stars around which disks are known to exist to at least 2
decades.

In an early stage of the pre-main sequence evolution, much of the stellar mass
is accreted onto the forming star through the disk, while angular momentum is
carried away. In a later phase, the stars are surrounded by a still
massive disk consisting of gas and dust, but the accretion rate is lower by
several orders of magnitude. It is during this phase that giant gas planets
and terrestrial planets (or at least "planetesimals", their kilometer sized
seeds) are thought to form. The disks dissipate on a time scale of several
Myr \cite{2001ApJ...553L.153H}, with the inner disk regions close to the star
being cleared first while the outer disk regions appear to survive somewhat
longer \citep[see e.g.][and references therein]{2001MNRAS.328..485C}.

During the early, high accretion phase the luminosity of the system is
dominated by the release of gravitational energy of accreting material and
hence the disks are called "active". Later, radiation from the stellar
photosphere is the main energy source; the disk absorbs part of the stellar
light and is thereby heated. The absorbed energy is re-emitted at infrared to
millimeter wavelengths, causing the "infrared excess" emission that is a key
feature of circumstellar material. Since the disks are not self-luminous during
this phase but merely reprocess stellar radiation, they are called
"passive". The passive disk phase lasts roughly an order of magnitude
longer than the earlier active disk phase, and passive disks are typically
much less enshrouded in natal cloud material than younger systems. Hence, most
observations of disks are of objects in the passive phase, this is true for
the observations reviewed here as well.

\subsection{Observations of circumstellar disks}
Observational studies of young stars and their disks are immensely numerous
and cover the electromagnetic spectrum from X-rays to radio wavelengths. In
the scope of this conference, it is useful to divide such studies into two
categories: those that do not \emph{spatially resolve} the disks, and those
that do. 

The majority of spatially unresolved studies focus on observing and modeling the
SED of young star+disk systems from the UV/optical to mm wavelengths.
Physical models of disks, and sometimes also envelopes, are employed to
reproduce the infrared excess emission \citep[see][and references
therein]{2007prpl.conf..555D}. Thereby, also the spatial structure of the
circumstellar environment is derived, albeit in an indirect way. High spectral
resolution optical and near-infrared spectroscopy of emission lines of mainly
Hydrogen that trace accretion activity often show strong day-to-day
variability, demonstrating that the innermost disk region is a highly
dynamical environment \citep[e.g.][and references
therein]{2007prpl.conf..479B}.  Spectro-polarimetry probes the structure of
gaseous inner accretion disks down to a scale of a few stellar radii
\cite{2005MNRAS.359.1049V}. These examples show that from spatially unresolved
observations, it is sometimes possible to derive properties of the
circumstellar material on scales much below the resolution limit of those
measurements, and even on scales that are still inaccessible even to the
highest spatial resolution observations available today. However, these
benefits often come at a price in the form of degeneracies in the derived
results that can be lifted by spatially resolved observations.

Since these are proceedings of a conference on high (angular) resolution
astronomy, the focus naturally is on observations that \emph{do} spatially
resolve the disk. Due to the small angular sizes of circumstellar disks,
interferometric techniques are commonly needed to resolve them. In some cases,
though, single telescopes may spatially resolve the outer disk regions of
nearby objects. Some disks have been resolved in optical/NIR scattered light,
after meticulous subtraction of the vastly dominant direct stellar light
\citep[e.g.][and references therein]{2007prpl.conf..523W}.  A handful of the
nearest circumstellar disks have been spatially resolved in infrared thermal
emission using single 4-10\,m class telescopes \citep[e.g.
][]{2004A&A...418..177V,2005prpl.conf.8417O,2007A&A...470..625D}

Intereferometric observations have spatially resolved circumstellar disks from
near-infrared to millimeter wavelengths. In millimeter continuum emission we
see the cold dust near the disk midplane and since the disks are usually
optically thin at this wavelength, the total dust mass can be estimated
\citep[e.g.][]{1994A&A...291..546H,2004A&A...416..179N}.  Millimeter emission
lines of mainly CO trace the disk kinematics on scales of $\sim$tens of AU,
and reveal keplerian velocity fields \cite{1997ApJ...490..792M}. In the
mid-infrared (by which here the 10\,$\mu$m atmospheric window is meant) we see
the disk surface and can probe the disk structure and dust properties on
scales $\sim$1-10\,AU. A truly vast amount of near-infrared interferometry has been
done in recent years. These observations are most sensitive to emission from
the inner edge
of the dusty disk, which is usually located at the radius where the
temperature equals the silicate dust sublimation temperature of
$\sim$1500\,K \citep[e.g.][]{2001ApJ...546..358M,2006ApJ...645L..77M,2005ApJ...622..440A,2005ApJ...635.1173A,2002ApJ...579..694M,2005ApJ...624..832M,2006ApJ...647..444M,2003ApJ...588..360E,2005ApJ...623..952E,2006A&A...451..951I}.
The inner edge of the dusty disk was proposed to have a locally increased scale height geometry \citep[][see also
figure~\ref{fig:leinert}]{2001ApJ...560..957D}, and indeed models that do include
such a "puffed-up" inner rim generally fit the near-infrared visibilities
considerably better than models that do not. Inward of the dust sublimation radius, evidence for hot gas is found in the
most recent spectrally resolved near-infrared measurements \citep[e.g.][]{2007ApJ...657..347E,2007Natur.447..562E}.

\vspace{0.2cm}

The observations reviewed here were done with MIDI \cite{2003SPIE.4838..893L},
the 10\,$\mu$m instrument of the VLT interferometer
\cite{2000SPIE.4006....2G}. MIDI is a 2-element interferometer, capable of
pairwise combination of the light from any of VLT's four 8.2\,m Unit
Telescopes and four 1.8\,m Auxiliary Telescopes. The spatial resolution of
MIDI is $\sim$5 to several tens of milli-arcseconds. For typical distances to
nearby star formation regions ($\sim$150-500~pc), this resolution is well
matched to the region where most 10\,$\mu$m emission is emitted:
$\sim$1-10\,AU from the central star. This is a highly interesting region,
since here terrestrial and giant gas planet formation take place, though this
process occurs near the mid-plane of the disk and is hidden from our view at
10\,$\mu$m. With MIDI we see only the disk \emph{surface}, and can study the
spatial structure of the disk as well as the dust properties in the surface
layer.

\subsection{Scope of this review}
Due to speaking time and writing space constraints, the scope of this
contribution is necessarily limited. Rather than attempting to give a complete
review of all work done with MIDI in the field to date, I will focus on a
number of selected studies that will serve to illustrate the possibilities and
limitations of the technique, and will give the reader a reasonable impression
of what has been achieved. The two main topics covered, disk structure and
dust mineralogy, are each illustrated with a few examples, and preluded by a
short general introduction.

\section{Disk structure on 1-10 AU scales}

\begin{figure}[t]
\includegraphics[height=16cm,angle=270]{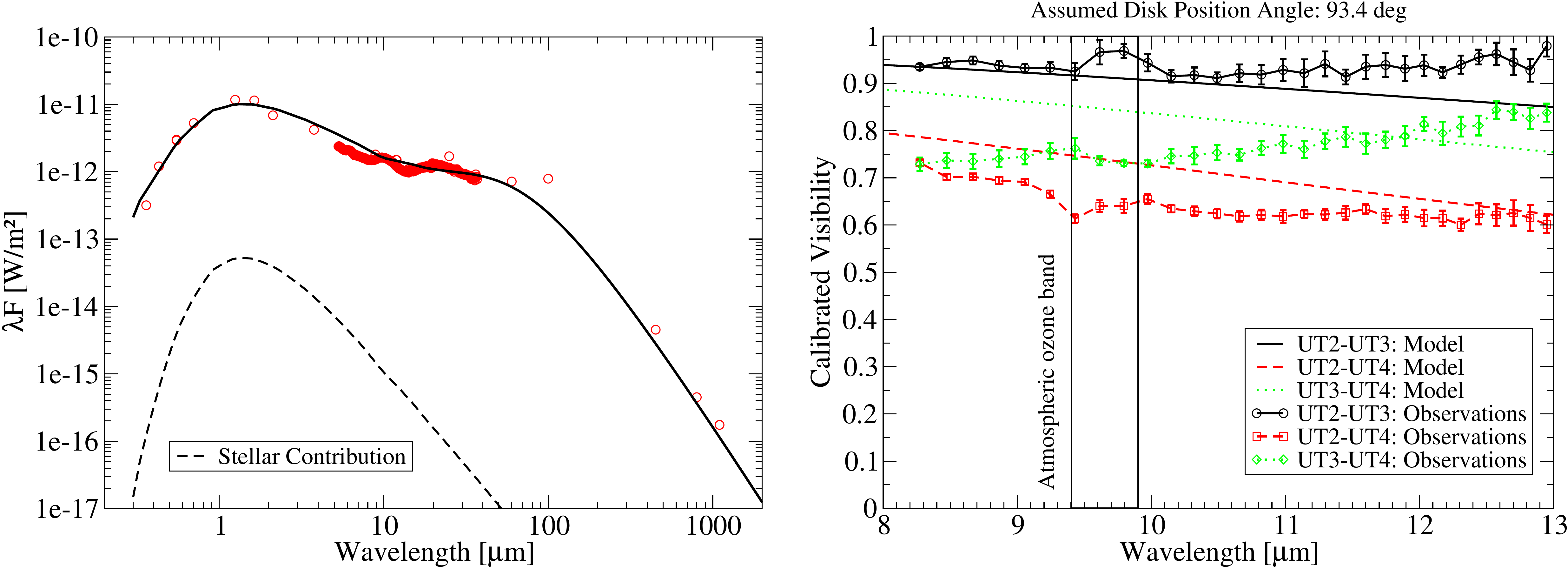}
\caption{\label{fig:quanz} Observations of FU~Orionis. Left: the SED from optical to millimeter wavelengths. Right: observed and modeled MIDI visibilities. The level of the observed visibilities indicates that the object is slightly but significantly resolved. The simple, parameterized disk model used to fit the observations reproduces the spatial extent of the emission well, but the deviation between the observed and modeled curves, in particular on the UT3-UT4 baseline, reveal that the actual source geometry is more complex than accounted for by the model. See section~\ref{sec:quanz} and Quanz et al. \cite{2006ApJ...648..472Q}.}
\end{figure}

\subsection{Introduction}
\label{sec:structure_intro}
Observing and modeling the spatial structure and dynamics of circumstellar
disks is a key topic in the study of star and planet formation. Long before
circumstellar disks could be spatially resolved, the observed infrared excess
was interpreted using initially simple and later ever more sophisticated disk
models. Disk models evolved from simple geometrically thin and optically thick
disks, through optically and geometrically thick flared disks
\cite{1987ApJ...323..714K} and refined versions of such models including an
optically thin disk atmosphere \cite{1997ApJ...490..368C,2001ApJ...560..957D},
to 2D and 3D radiative transfer disk models in which the temperature and
vertical structure are computed self-consistently \cite{2004A&A...417..159D}.
Physical disk models remain an essential tool to interpret also spatially
resolved data.

Since circumstellar disks contain material at a vast range of temperatures,
observations at a range of wavelengths are required to characterize their
emission and study their structure. On sub-AU scales we find very hot material
that is best seen in the near-infrared, whereas material near the disk
mid-plane and in the cold outer disk regions is best studied at millimeter
wavelengths. At 10\,$\mu$m we probe emission from the disk surface at scales
of $\lesssim$20\,AU. A key study into the structure of this region of the disk
was that of Meeus et al. \cite{2001A&A...365..476M}. While analyzing ISO
spectra of a sample of Herbig~Ae stars, they found that the infrared SEDs of
these objects come in two flavors: stars with approximately a power law
spectrum throughout the infrared, and stars that \emph{additionally} display a
relatively cool component peaking around 60\,$\mu$m. The objects with the cool
component were dubbed "group~I", the ones lacking the cool component
"group~II". Meeus et al. speculated that stars showing group~I SEDs exhibit
flared disks, whereas group~II objects have "flat", or "self-shadowed" disks
(see figure~\ref{fig:leinert}), yet their observations had vastly insufficient
spatial resolution to directly assess the disk geometry.

In this section we will see the results of MIDI studies of an actively
accreting disk, a sample of presumably more evolved passive disks, and an even
more mature system that is thought to be in the transition phase towards a
gas-deprived debris disk.

\subsection{FU~Orionis: an active accretion disk}
\label{sec:quanz}

FU~Orionis is the prototype FUOR variable. These objects show "outbursts"
during which the system may brighten by $>$5\,mag at optical wavelengths over
the course of several months, followed by a slow decline which can last tens
of years or even centuries. It is thought that the brightening is caused by an
episode of strongly enhanced accretion, induced by a yet not fully understood
instability in the inner disk region. During an outburst, the system
luminosity is fully dominated by the release of gravitational potential energy
of the accreting material. See Hartmann \& Kenyon \cite{1996ARA&A..34..207H}
for a detailed overview of the FUOR phenomenon.

Figure~\ref{fig:quanz} shows the SED and MIDI observations of FU~Ori as
presented by Quanz et al. \cite{2006ApJ...648..472Q}. These observations
spatially resolve the FU~Ori disk, for the first time at 10\,$\mu$m.  The
authors model the SED and MIDI visibilities simultaneously, and find that a
disk model with a broken power law temperature profile provides the best fit
to their observations. Within $\sim$3\,AU from the central star, the
temperature follows $T$\,$\propto$\,$R^{-0.75}$, whereas at larger radii,
$T$\,$\propto$\,$R^{-0.53}$. These results indicate that the heating in the
inner disk region is dominated by the release of gravitational energy (which
has a radial dependence of d$E$/d$t$\,$\propto$\,$R^{-3}$, per unit area), and
in the outer region by the absorption of radiation (which follows an $R^{-2}$
law in the optically thin disk atmosphere). Thus, in the inner region the disk
has a temperature profile typical of an active accretion disk, whereas further
out the temperature equals that of a passive, reprocessing disk. Such behavior
can be understood: even in an accretion dominated system, the main heating
source of disk material at larger radii will be absorption of radiation
emitted in the central region, due to the different radial dependencies of
heating by accretion and absorption. Whether the "central engine" is a stellar
photosphere like in a "standard" T\,Tauri system, or a luminous inner
accretion disk as in a FUOR object in outburst, is irrelevant in this respect.

\subsection{HAe stars: the structure of passive disks}
\label{sec:leinert}

\begin{figure}[t]
\includegraphics[height=16cm,angle=270]{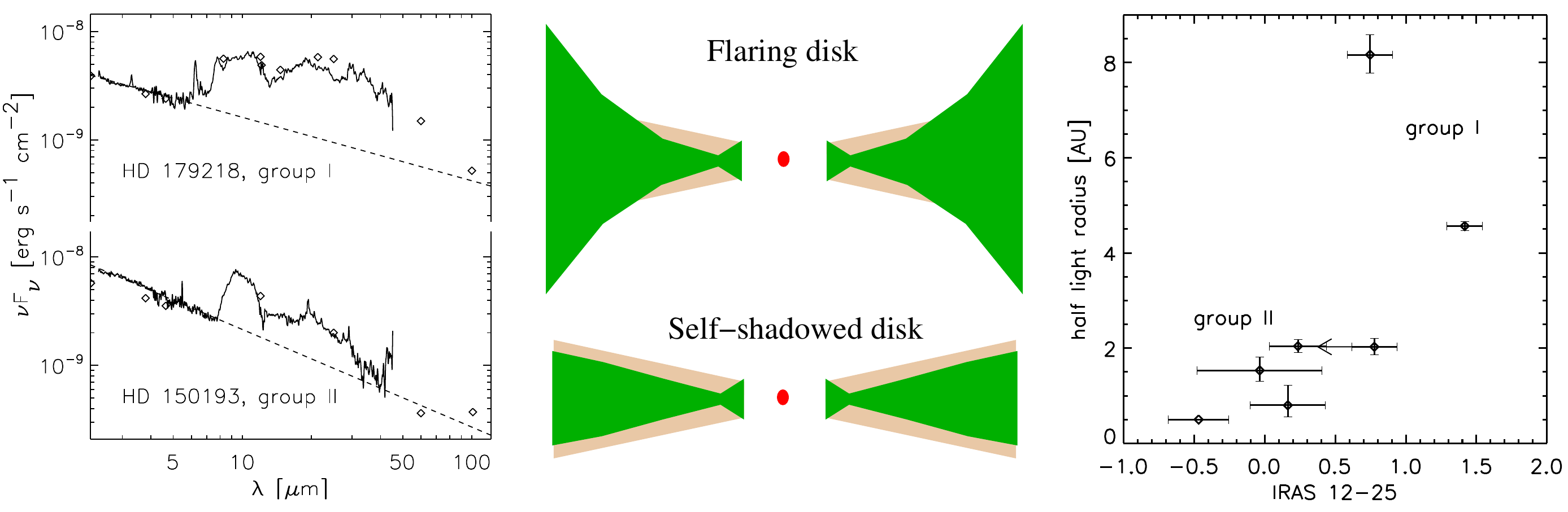}
\caption{\label{fig:leinert} Spectral energy distributions, qualitative disk
  models and MIDI observations of HAe stars.  {\it Left}: typical SEDs of a
  group~I and a group~II object. {\it Middle}: disk geometries proposed to
  explain the difference in IR spectral shape. {\it Right}: the spatial extent
  of the 10\,$\mu$m emission as measured with MIDI. The group~II sources
  appear spatially more compact than the group~I sources, confirming the
  SED-based hypothesis. See section~\ref{sec:leinert} and Leinert et al.
  \citep[][]{2004A&A...423..537L}.  }
\end{figure}

As found by Meeus et al. \cite{2001A&A...365..476M}, the infrared SEDs of circumstellar disks show a range of spectral shapes,  that were proposed to be related to the spatial structure of the disks. Disks with "red" IR colors were thought to have flared shapes, whereas disks with "blue" colors would exhibit a flat (or "self-shadowed) geometry. This hypothesis makes clear predictions on the spatial extent, or "size", of the 10\,$\mu$m emission in circumstellar disks: sources with red SEDs (Meeus group~I) should appear larger on the sky, whereas blue sources (group~II) should show more compact emission (see also section~\ref{sec:structure_intro}).

In order to test this hypothesis, Leinert et al. \cite{2004A&A...423..537L} used MIDI to observe a sample of nearby HAe stars including both group~I and group~II objects. All observed disks were spatially resolved at 10\,$\mu$m, and the spatial extent of the emission could be estimated. The results are shown in the right panel of figure~\ref{fig:leinert}, in which the "size" of the emission\footnote{It is not straightforward to define the "size" of the emission from a circumstellar disk at 10\,$\mu$m. Disks emit at a fairly large range of radii at this wavelength, and the radial intensity profile cannot be described by a single parameter (see \cite{2005A&A...441..563V} for a detailed discussion). Leinert et al. define the size as follows: they fit a model of an optically thin dust disk around each individual star to match the observed MIDI visibilities. The "size" of the emission is then defined as the radius within which half of the 10\,$\mu$m emission is emitted.} is displayed against the IRAS 12-25~$\mu$m color\footnote{In one of the group~II stars, HD~142527 located at (0.8,2.0) in the rightmost panel in figure~\ref{fig:leinert}, the IRAS~25~$\mu$m flux is heavily contaminated by extended emission not directly related to the disk itself. If this is corrected for, this object moves to position (0.4,2.0) in the diagram as indicated by the arrow.}. The two group~I sources clearly have spatially more extended emission than the group~II sources in the sample. Thus, the first observations that directly spatially resolve the disk on the relevant scales qualitatively confirm the SED-based hypothesis of Meeus et al. 

\subsection{TW~Hya, a transition disk}
\label{sec:ratzka}

\begin{figure}[t]
\includegraphics[height=16cm,angle=270]{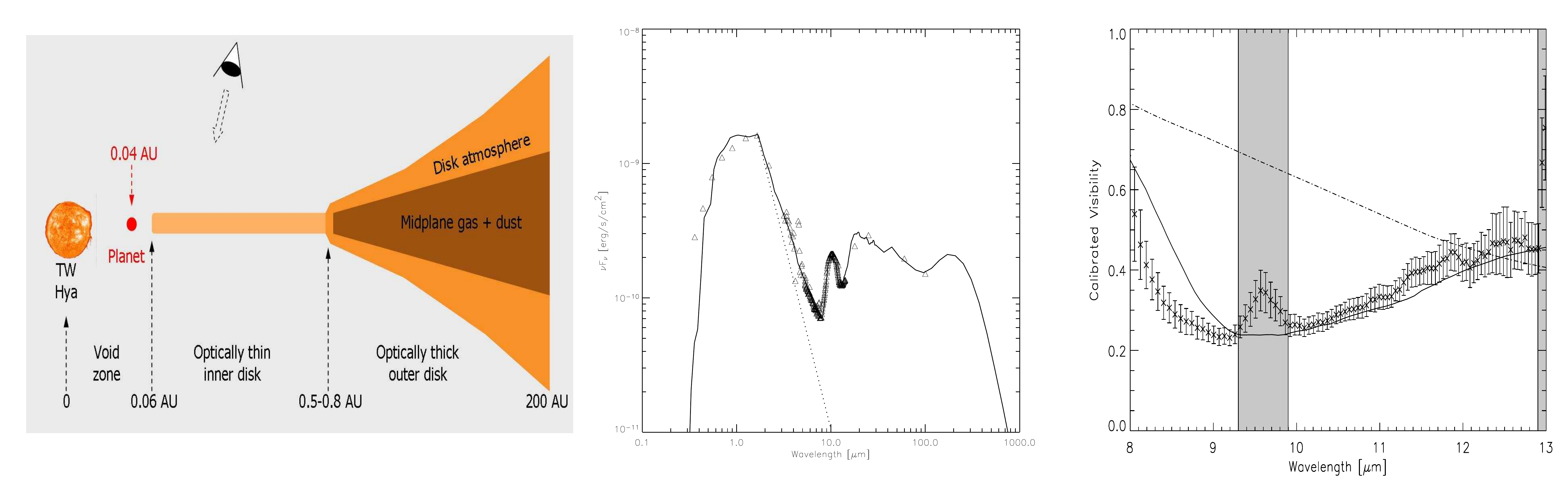}
\caption{\label{fig:ratzka}
  The transition disk system TW~Hya. {\it Left}: a sketch of the system. {\it
    Middle}: the observed SED and a radiative transfer disk model fit, that
  was simultaneously fitted to the MIDI visibilities as well. Units are
    wavelength in micrometer on the horizontal axis running from 0.1\,$\mu$m to
    1000\,$\mu$m on a logarithmic scale, and flux ($\nu F_{\nu}$) on the
    vertical axis running from 10$^{-11}$ to 10$^{-8}$
    erg\,s$^{-1}$cm$^{-2}$ on a logarithmic scale. {\it Right}:
  the observed and modeled MIDI visibilities. Also indicated with a
  dashed-dotted line is a previous model, constructed before the MIDI
  observations were available.  See section~\ref{sec:ratzka} and Ratzka et al.
  \citep[][]{2007A&A...471..173R}.}
\end{figure}

Transition disk systems are thought to be in an evolutionary stage between
passive, gas-rich disk phase and the debris disk phase. The dissipation of the
disk is on-going in these objects. Transition disks are characterized by
strongly reduced IR excess emission at near-infrared wavelengths compared to
younger systems but a still strong IR excess at wavelengths longward  of
$\sim$10\,$\mu$m. This indicates that the innermost disk regions have already
been largely cleared of material whereas the outer disk regions are still
relatively intact. What physical mechanism dominates the disk dissipation
process is not yet clear, the roles of photo-evaporation, accretion and planet
formation need further investigation \citep[e.g.][]{2005ApJ...629..881H,2007ApJ...670L.135E,2006AJ....132.2135S}.

In this respect, TW~Hya is a particularly interesting system: recently a giant
gas planet was found orbiting this star \cite{2008Natur.451...38S}. The orbit
of the planet lies just within the inner radius of the disk as measured using
2\,$\mu$m interferometry \cite{2006ApJ...637L.133E}, and it is the first
planet found orbiting such a young star still surrounded by a disk. TW~Hya is
an estimated 10~Myrs old and at a distance of 51\,pc it is  very nearby,
making it an excellent target for spatially resolved studies \cite{1999ApJ...512L..63W}. 

Figure~\ref{fig:ratzka} shows a sketch of the system, as well as the SED and
MIDI observations discussed here. The SED shows a small but significant IR
excess above the photospheric level in the 3-8\,$\mu$m range, indicating small
amounts of hot material are present close to the star. Calvet et
al. \cite{2002ApJ...568.1008C} model the SED with an optically thin gas and
dust distribution close to the central star, surrounded by an optically thick
disk. They put the edge of the optically thick disk around 3-4\,AU, where it
does not contribute to the 10\,$\mu$m emission; in their model, the 10\,$\mu$m
emission arises solely in the optically thin inner disk region. Hughes et
al. \cite{2007ApJ...664..536H} present millimeter interferometry of the
continuum emission probing the bulk dust in the TW\,Hya disk. They find the
emission has a central depression, indicating an inner gap of similar size to
that modeled by Calvet et al. \cite{2002ApJ...568.1008C}.

Ratzka et al. \cite{2007A&A...471..173R} observed TW~Hya with MIDI. Their
observations are shown in the right panel of figure~\ref{fig:ratzka}. They
model the SED and the MIDI observations using a qualitatively identical disk
configuration to that of Calvet et al., yet the MIDI observations force the
transition region between the optically thin inner disk and the optically
thick outer part to be much closer to the central star, at $\sim$0.6-1.0\,AU. In the
Ratzka model, the 10\,$\mu$m emission is dominated by the inner edge of the
optically thick outer disk. The 10\,$\mu$m visibilities predicted by the
Calvet and Ratzka models are shown in figure~\ref{fig:ratzka} with a dashed
and solid curve, respectively.  While both models reproduce the visibilites
measured at 2\,$\mu$m \cite{2006ApJ...637L.133E}, only the Ratzka
model can account for the 10\,$\mu$m emission. Thus, TW\,Hya
illustrates both the importance of having multi-wavelength observations, as well as the need for spatially resolved observations to raise degeneracies that may occur when the disk structure is derived solely from the SED.

While the MIDI observations convincingly place the "transition region" to the
optically thick disk region at $\lesssim$1\,AU, the central "gap" in the
mm~emission is clearly larger \cite{2007ApJ...664..536H}. These observations
probe the bulk of the dust mass, located close to the disk mid-plane.
Possibly, most of the dusty material in the 1-4\,AU region has already been
removed (or trapped in larger bodies that are not seen), while there is still
enough dust present to make the region optically thick at 10\,$\mu$m; this
would require only a tiny fraction of the material needed to make the disk
optically thick at mm wavelengths.  If true, this may argue for a dissipation
mechanism "from within the disk" such as accretion or planet formation, as
opposed to a mechanism acting "from outside" (i.e. photo-evaporation).

\section{Dust mineralogy on 1-10 AU scales}
\label{sec:mineralogy}

\subsection{Introduction}
\label{sec:mineralogy_intro}

Dust is a fundamental constituent of circumstellar disks. Even though it
contains only $\sim$1\% of the disk mass, it is the dominant source of opacity
in the disk. Therefore, both radiative heating and cooling of disk material,
and thereby the disk temperature, are goverened by the dust properties -
except very close to the central star where it is too hot for dust to survive
(T$\gtrsim$1500\,K). The disk temperature in turn governs intra-disk
chemistry, the location where important species such as H$_2$O and CO freeze
out, and the (vertical) structure of the disk in a balance between gas
pressure and the gravitational field \citep[see e.g.][and references
therein]{2007prpl.conf..555D}. Moreover, dust grains may collide with other
dust grains and stick, initiating a chain of growth that eventually leads to
kilometer sized seeds of terrestrial planets, and of the rocky cores of giant
gas planets \citep[see][and references
therein]{2007prpl.conf..783D,2007Natur.448.1022J}. Dust, in short, is
important stuff. See Natta et al. \cite{2007prpl.conf..767N} for an overview
of dust properties in disks around young stars.

\begin{figure}[t]
\begin{tabular}{lr}
\begin{minipage}[h]{10cm}

\hspace{-0.4cm}
\includegraphics[width=10cm,height=5cm,angle=0]{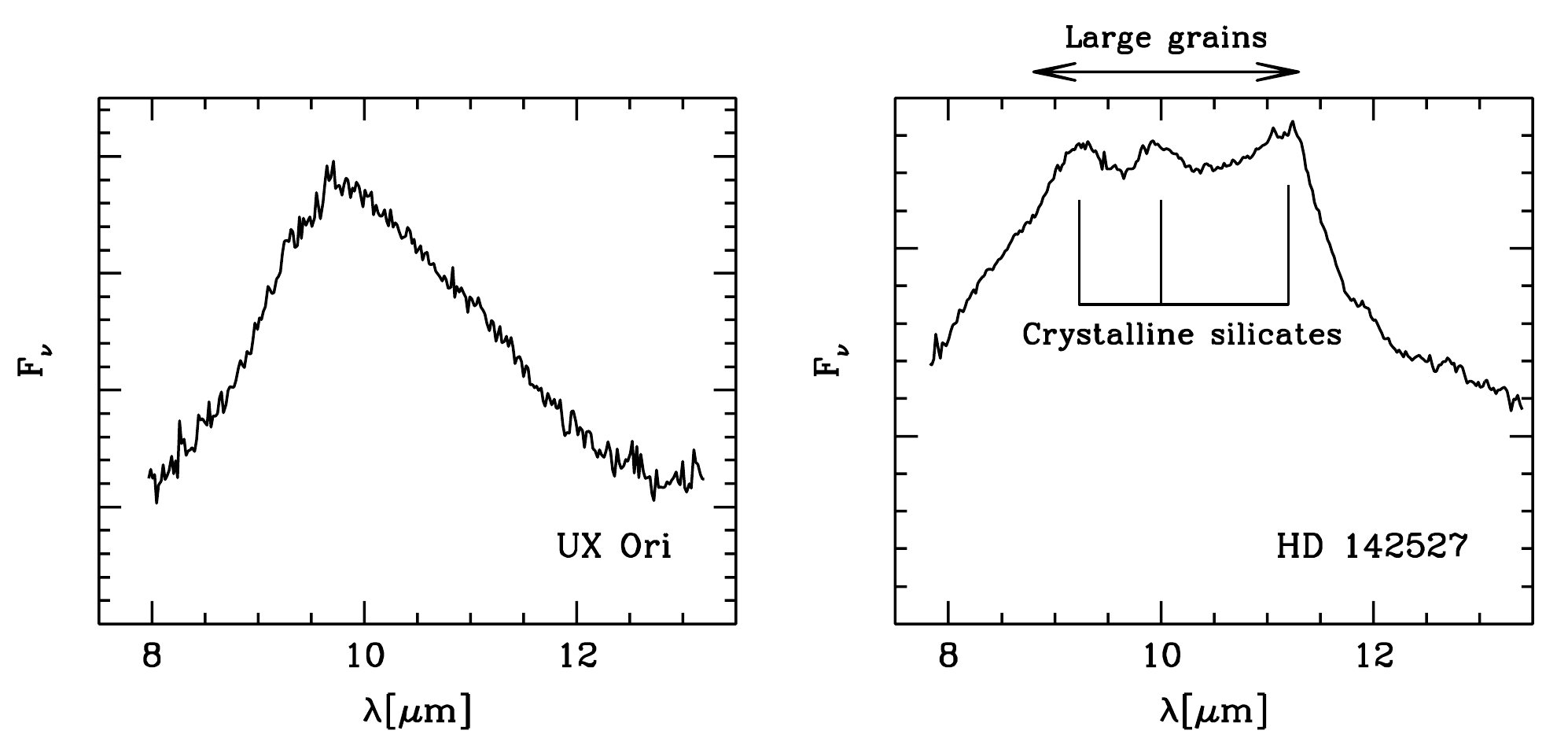}
\end{minipage}
&
\begin{minipage}[h]{5.3cm}\caption{\label{fig:processing}
The effects of dust processing on the 10\,$\mu$m silicate feature. The left
spectrum is typical of "pristine" dust as found in the ISM: small, amorphous grains. In the right spectrum the broader, flat-topped silicate band implies
grain growth, whereas the additional narrower bands witness the presence of
crystalline material.}
\end{minipage}
\end{tabular}
\end{figure}

The dust in molecular clouds and circumstellar disks consists mainly of carbon
and silicates of various stoichiometries and lattice structures. Whereas the
opacity of carbon does not show much spectral structure in the infrared,
silicate grains have strong resonances that can be used to derive the
composition, size, and lattice structure of the dust grains. Dust in the
ISM consists of small grains in which the molecular building blocks are
randomnly ordered, i.e. the material is amorphous \cite{2004ApJ...609..826K}. In circumstellar disks, as
well as in solar system comets which are thought to be frozen records of the
disk that once surrounded the young sun, the dust is distinctly different in
two respects. First, the dust grains are often larger than the typical ISM
value of $\sim$0.1\,$\mu$m \citep[e.g.][]{2003A&A...400L..21V,2007ApJ...659.1637S}. Second, a significant fraction of disk and comet
material may be in a \emph{crystalline} form
\citep[e.g.][]{1997EM&P...79..265W,1997EM&P...79..247H,2001A&A...375..950B,2005A&A...437..189V,2005ApJ...622..404K}. Figure~\ref{fig:processing}
illustrates how the "10 micron silicate feature" can be used as a
diagnostic tracing the growth of dust grains from sub-micron sizes to several
micron, and the conversion of amorphous to crystalline silicates.

In particular crystallization of the initially amorphous ISM dust is a highly
interesting process. Contrary to growth, which may occur anywhere in the
relatively high density environment of a circumstellar disk, crystallization
requires rather "special" circumstances: high temperatures (T$\gtrsim$900\,K).
Such temperatures surely prevail close to the central star, yet crystals are
also found at much lower temperatures in some disks, and are a common
ingredient of solar system comets that formed in low temperature
($\lesssim$150\,K) regions in the solar nebula. How crystals came to be
present in these cold regions is hotly debated, rivaling theories holding that
they were formed there in-situ during transient heating events induced by
shocks \cite{2002ApJ...565L.109H} or electric discharges \cite{1998A&A...331..121P}, or that they were created in the hot inner disk
regions and transported outward via radial mixing \cite{2004A&A...413..571G}. Thus, crystalline silicates
are potentially a powerful tracer of turbulence and large scale mixing
processes in disks. Therefore, it is interesting not only to know how much
crystalline dust is present in a disk, but also \emph{where} it is located.

\vspace{0.1cm}

With its spectroscopic capabilities and high angular resolution, MIDI has the
unprecedented capability of probing the properties of the dust particles in
the disk surface on scales of 1-10\,AU, typically. Yet, a word of caution is
appropriate here. Rather than making true images of sources, MIDI measures
"visibilities": discrete components of the fourier transform of the image.
Measurements on long baselines probe emission on smaller scales than those on
short baselines, yet this emission is not directly located\footnote{In a
  2-element optical interferometer, such as MIDI, only the amplitude of the
  (complex) visibility can be measured; the phase is lost due to the
  disturbing effect of the Earth atmosphere. This prohibits true image
  reconstruction such as routinely done e.g. at radio wavelengths (aperture
  synthesis imaging), and makes the interferometer insensitive to the exact
  location of the source (but fortunately not to the shape of the emission).}.
A commonly applied strategy is to compare the "correlated flux" (the amount of
signal that is coherent on the respective baseline) to the "total flux" as
measured by MIDI. The correlated flux arises from a small region, typically
less than a few~AU in diameter, and is attributed to the innermost disk
regions. This approach is justified as long as the correlated flux is a
significant fraction of the total flux: only in the innermost disk region we
may reasonably expect so much flux  to be emitted in such a small
region. However, if the ratio of corrrelated and total flux (by definition
equal to the visibility) gets close to 0, this approach is no longer valid and
the spectrum in correlated flux can not be directly attributed to the inner
disk; more detailed modeling is then needed.

That being said, let us now compare some MIDI measurements of circumstellar
disks with the representative spectra of "pristine, ISM like" dust and of
"processed, comet-like" dust displayed in figure~\ref{fig:processing}! We will
see some examples of MIDI studies of young $\sim$solar type stars (T Tauri
stars) as well as slightly more massive HAe stars. Note, though, that in terms
of disk properties T\,Tauri and HAe stars are very similar and that the
division may be somewhat artificial. In practical terms, HAe stars are
generally somewhat easier to observe due to their higher 10\,$\mu$m
brightness.

\subsection{Mineralogy in Herbig Ae star disks}
\label{sec:vboekel}

\begin{figure}[t]
\includegraphics[width=16cm,height=7cm,angle=0]{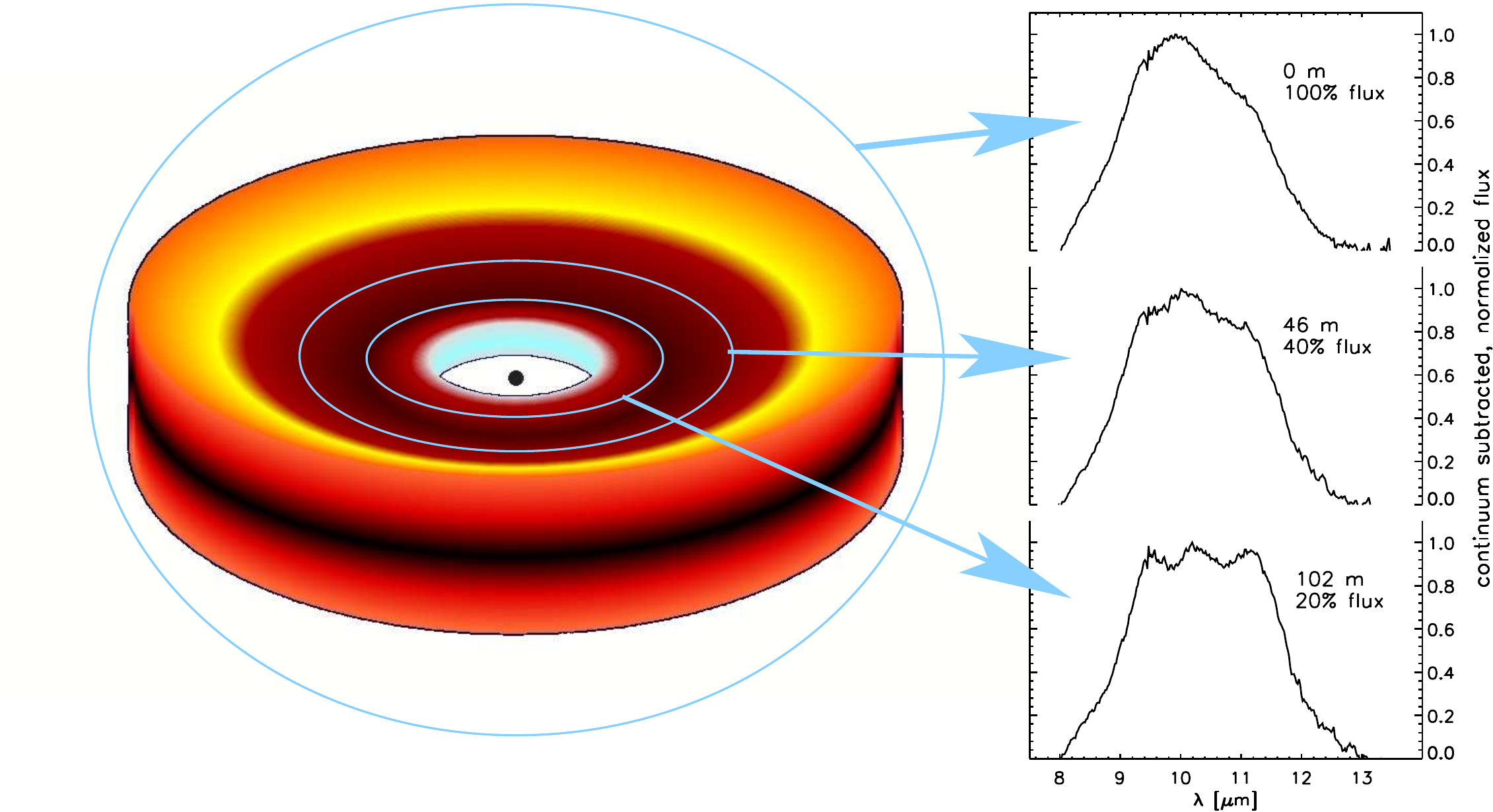}
\caption{\label{fig:vboekel}
Continuum subtracted 10\,$\mu$m silicate emission spectra of Herbig~Ae star
HD\,144432, taken with MIDI on different baselines. The upper spectrum
shows the total emission, arising in the entire disk region warm enough to
emit at 10\,$\mu$m (i.e. at $R$\,$\lesssim$\,20~AU). The middle spectrum
represents the correlated flux measured at a baseline of 46\,m, which is
dominated by emission from the disk region at $R$\,$\lesssim$\,3~AU. At the
bottom, the correlated flux at a baseline of 102\,m, representative of the
disk region at $R$\,$\lesssim$\,1.5~AU is shown. As witnessed by the shape of
these spectra, the dust properties in the disk surface strongly depend on the
distance to the central star. See section~\ref{sec:vboekel} and \citep[][]{2004Natur.432..479V}.
}
\end{figure}

In figure~\ref{fig:vboekel} MIDI spectra of the dust in the disk around
Herbig~Ae star HD\,144432 are shown that clearly reveal the 10\,$\mu$m silicate
feature \cite{2004Natur.432..479V}. The upper spectrum shows the integrated
flux as measured by a single telescope (one may consider it the correlated spectrum at a baseline of 0\,m). Here, the source is spatially unresolved and we see the light from the entire disk region that is warm enough to emit at 10\,$\mu$m, i.e. roughly the central 10-20\,AU. Comparing to the spectra in figure~\ref{fig:processing} we see that most of the material visible here is pristine, but the small ``shoulder'' at 11.3\,$\mu$m indicates that there is some processed material \emph{somewhere} in the system. The middle and lower plot show the correlated spectrum as measured by MIDI on baselines of 46 and 102\,m. The emission seen here is dominated by the disk regions within 3 and 1.5\,AU of the central star, respectively. The correlated spectrum at a baseline of 102\,m, which probes the smallest spatial scales, is very similar to the ``evolved'' spectrum in figure~\ref{fig:processing}. Compositional fits to these spectra were made, including both amorphous and crystalline silicates with grain sizes of 0.1 and 1.5\,$\mu$m as dust species, in order to study both grain growth and crystallization. The fraction of crystalline silicates was found to be approximately 5, 12 and 36 percent for the spectra at baselines of 0, 46 and 102\,m, respectively. The fraction of material contained in large grains is 40, 86 and 93 percent by mass. Thus a clear trend is seen, indicative of both crystallinity and average grain size in the surface layer of the disk decreasing with distance to the central star.

Similar results were obtained for two additional objects (HD\,163296 and
HD\,142527). All three disks display a much higher degree of processing in
their innermost regions than further out. In the case of HD\,142527, there are
crystalline silicates present also at larger radii. Also, a chemical gradient
in the composition of the crystals is seen, with a forsterite dominated
spectrum closest to the star, and more enstatite at larger radii. These data
support the radial mixing scenario for the origin of crystalline siliacetes

\subsection{Mineralogy in T Tauri star disks}
\label{sec:mineralogy_TTau}

The young star RY~Tau, lies at a distance of $\sim$140\,pc and has an
estimated mass and age of 1.7\,M$_{\odot}$ and 7\,Myr
\cite{1999A&A...342..480S}, respectively. Its circumstellar disk is oriented
relatively close to edge-on from our vantage point
($i$\,$\lesssim$\,70$^{\circ}$, \cite{2008A&A...478..779S}) and the object
shows significant optical variability. Its mass puts it roughly at the border
between the T~Tauri and Herbig~Ae regime (but bear in mind that in terms of
disk properties this division may be largely artificial).

\begin{figure}[t]
\begin{tabular}{lr}
\begin{minipage}[h]{10cm}
\includegraphics[width=9cm,angle=0]{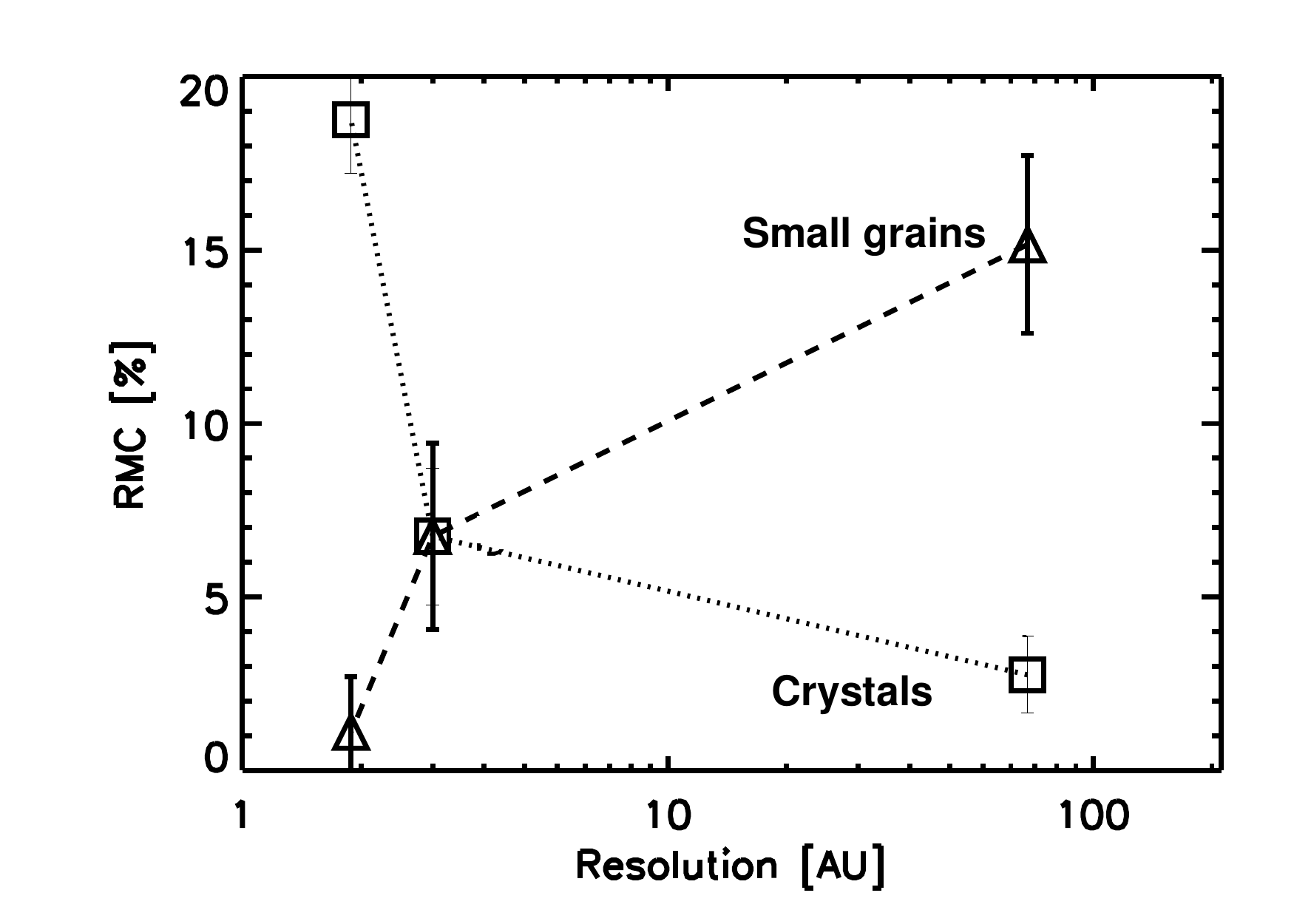}
\end{minipage}
&
\begin{minipage}[h]{5cm}\caption{\label{fig:schegerer} The spatially resolved
    mineralogy of T~Tauri star RY~Tau obtained with MIDI. Shown are the
    abundances of crystalline silicates and small grains as a function of the
    spatial scale probed. As the intereferometer "zooms" in further, the
    abundance of crystals increases and that of small grains decreases. See
    section~\ref{sec:mineralogy_TTau} and Schegerer et al. [65].}

\end{minipage}
\end{tabular}
\end{figure}

RY~Tau was observed with MIDI by Schegerer et al. \cite{2008A&A...478..779S},
who study the disk structure (not discussed here) as well as the mineralogy of
this object (parameters). As shown in figure~\ref{fig:schegerer}, they find 
strongly radially dependent dust properties. As the interferometric baseline
is increased and emission on smaller scales is probed, the abundance of
crystalline silicates increases dramatically. The abundance of small grains
decreases strongly close to the central star, indicating that grain growth has
proceeded furthest in the central disk regions. This behavior exactly mimics
that observed in HAe star disks.

\vspace{0.2cm}

Let us now take a look at the mineralogy of a disk around a star of sub-solar
mass: the transition disk TW~Hya, whose disk structure was discussed in
\cite{2007A&A...471..173R} and section~\ref{sec:ratzka}. The mass and age of
this object are estimated to be 0.6\,M$_{\odot}$ and 5-15\,Myr, respectively \cite{1999ApJ...512L..63W}. 

\begin{figure}[t]
\includegraphics[width=16cm,height=5.5cm]{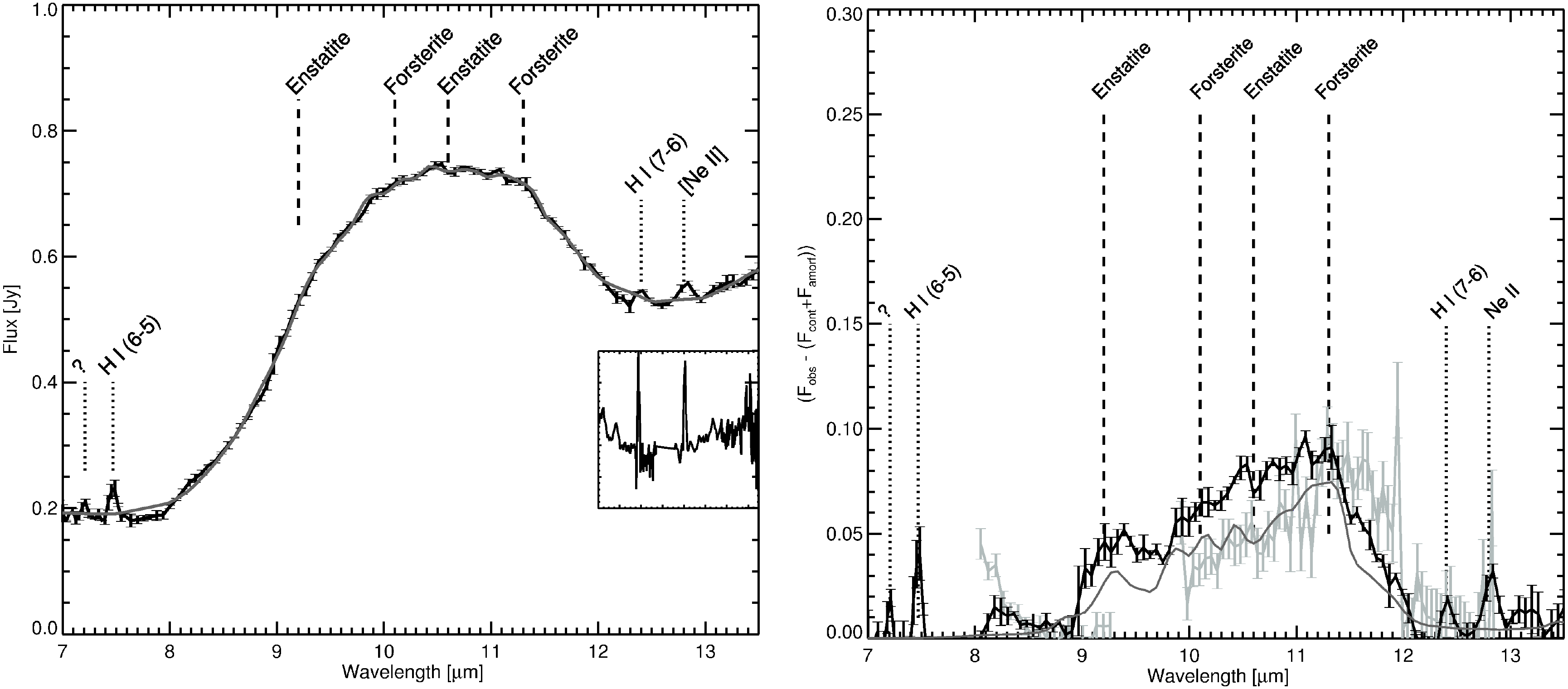}
\caption{\label{fig:ratzka_mineralogy}
10\,$\mu$m spectroscopy of the transition disk system TW~Hya. {\it Left}: the
Spitzer IRS spectrum. Wavelengths of spectral features of crystalline dust
species as well as gas lines are indicated. {\it Right}: the MIDI correlated
flux spectrum (light grey curve) as well as a dust model fit to this spectrum
(dark grey curve). Also shown is the residual Spitzer IRS spectrum after
subtraction of the best fit continuum and amorphous dust model. These data
show that the crystalline dust is strongly concentrated in the innermost disk
region. See section~\ref{sec:mineralogy_TTau} and Ratzka et al. \cite{2007A&A...471..173R}
}
\end{figure}

Figure~\ref{fig:ratzka_mineralogy} shows spectroscopy of the TW~Hya disk in the 10\,$\mu$m region. In the left panel we see the spatially unresolved Spitzer spectrum, showing a strong silicate emission feature. A compositional analysis reveals that the dust consists mostly of amorphous silicates that have undergone a significant amount of grain growth compared to ISM sizes. Small amounts of crystalline material are also detected, but not located due to insufficient spatial resolution. In the right panel of figure~\ref{fig:ratzka_mineralogy} we see the continuum-subtracted MIDI spectrum in correlated flux, that traces emission on sub-AU scales (light grey curve) as well as dust model fit to this spectrum (dark grey curve). Also shown is the Spitzer spectrum, after subtraction of the best fit continuum and amorphous dust model (black curve). This spectrum is dominated by emission from crystalline silicates, and is remarkably similar to the MIDI correlated flux spectrum. Once more, as was the case in the previous examples of stars of 1-2\,M$_{\odot}$, we find that the crystalline silicates are concentrated in the innermost disk region.

\section{Summary}
In this contribution, 10\,$\mu$m interferometric observations of the disks
around young stars performed with MIDI were reviewed. This overview was by no
means complete, but rather intended to give the reader an idea of the
capabilities of MIDI. We saw how the spatially resolved observations were used
to test hypotheses on disk structure based on SED modeling, generally finding
good agreement but occasionally finding discrepancies that reveal degeneracies
occuring if the disk structure is derived solely from the SED. The properties
of the dust in the surface layer of the disks were found to depend strongly on
the location within the disk: close to the central star the material is highly
crystalline, whereas at larger radii it is mostly amorphous. The size of the
dust grains is also larger in the inner disk regions than further out.

\bibliographystyle{iopart-num}
\bibliography{references}

\end{document}